\documentclass[twocolumn,superscriptaddress,floatfix,prl,showpacs]{revtex4}
\usepackage{graphicx}
\usepackage{color}

\begin{document}
\title{Hall effect in ferromagnetic nanomagnets: magnetic field dependence as an evidence of
inverse spin Hall effect contribution}
\author{Vadym Zayets}
\affiliation{National Institute of Advanced Industrial Science and Technology (AIST), Umezono 1-1-1, Tsukuba, Ibaraki, Japan}
\author{Andrey S. Mishchenko}
\affiliation{RIKEN Center for Emergent Matter Science (CEMS),  
2-1 Hirosawa, Wako, Saitama, 351-0198, Japan}

\begin{abstract}
We measure magnetic field dependence of the Hall angle in a metallic ferromagnetic nanomagnet with stable
local magnetic moments where the adopted mechanisms of Hall effect predict linear plus a constant
dependence on the external field originating from the ordinary and anomalous Hall effects, respectively. 
We suggest that the experimentally observed deviations from this dependence is caused by the  
inverse spin Hall effect (ISHE) and develop a phenomenological theory, which predicts a unique nonlinear 
dependence of the ISHE contribution on the external magnetic field.
Perfect agreement between theory and experiment supports the considerable role
of the ISHE in the Hall transport in ferromagnetic metals.
\end{abstract}

\maketitle

The Hall effect (HE) desribes a generation of an electric current perpendicularly to the bias current, which flows along  an applied electric field.
The magnitude of the HE is  measured by the Hall angle 
$\alpha_{\mbox{\scriptsize HE}}=\sigma_{xy}/\sigma_{xx}$, which is defined as the  ratio of nondiagonal 
$\sigma_{xy}$ and diagonal $\sigma_{xx}$ conductivities.
Early experiments of a measurement of the  HE in a nonmagnetic metal and semiconductor detected the simplest counterpart 
of the phenomenon, which is called the ordinary Hall effect (OHE) \cite{Hall1879}.  
Further studies of the HE in magnetic materials revealed several other mechanisms 
each of which is fundamentally different from the OHE. 

Historically, the first considered mechanism of the HE is the OHE, in which the 
carrier motion is perpendicular to the external magnetic fileld and the bias current is generated by the Lorentz force.
The standard OHE contribution to the Hall angle is proportional to the external 
magnetic field $\sim \alpha_{\mbox{\scriptsize OHE}} H$, where 
$\alpha_{\mbox{\scriptsize OHE}}$ is the OHE coefficient.
Another HE contribution, which exists in a ferromagnetic material, is the Anomalous Hall effect (AHE). The AHE  
occurs due to the scattering of carriers on the aligned local magnetic moments in a ferromagnet and  therefore the AHE
is proportional to the local magnetic moments or the magnetization $M$ \cite{PughRostoker1953}. 
The AHE contribution to the Hall angle  is independent on the external magnetic field $H$ 
if the local magnetic moments are field- independent.
The joint contribution of the OHE and AHE to the Hall angle is the sum of 
two contributions: the field- independent 
AHE $\sim \alpha_{\mbox{\scriptsize AHE}}$  and the linear OHE  $\sim \alpha_{\mbox{\scriptsize OHE}} H$.
The above constant plus linear dependence, which is the standard case encountered in the plenty 
of magnetic compounds, is a target of the recent theoretical efforts \cite{Sinova}.
 
More complicated dependence of the HE angle on the external magnetic field, which does not fit into constant plus linear dependence,
points either to the dependence of the localized magnetic moment on the external magnetic filed or to the
importance of an additional mechanism of HE which is intrinsically sensitive to the external magnetic field $H$.  
One possible candidate for such a mechanism is the inverse spin Hall effect (ISHE), which describes the fact that an electrical current is created perpendicularly to the flow of the spin- polarized current  \cite{Saitoh,Tinkham, Maekawa}.  The ISHE originates from  the spin- dependence of electron scatterings due to the Spin-Orbit (SO) interaction.  For example, in case of a spin- polarized electron gas, the scattering probabilities of spin- up and spin- down electrons are different for the left and right scattering directions. Since there are more electron that are scattered e.g. to the left than to the right, an electron current flows from the left to the right.  This mechanism of the ISHE is called the skew scattering mechanism  \cite{r30}. The ISHE requires the existance of the spin-polarized electrons and therefore it occurs only in a  spin- polarized electron gas and is proportional to the number of the spin- polarized electrons. In a ferromagnetic metal the electron gas is spin- polarized even in an equlibrium and any electrical current in a ferromagnetic metal is spin- polarized. As a result, any electrical current in a ferromagnetic metal always experiences the ISHE and  there is a flow of the Hall current perperpendiculary to the bias current.  The ISHE- type Hall current is proportional to the  number of spin polarized electrons in the ferromagnetic metal.
There are several methods to change the number of spin- polarized electron in a ferromagnetic metal and therefore the magnetitude of the ISHE. E.g. the spin polarization of the conduction electrons can arise due to the influence of an external 
magnetic field, the mechanism  considered by Landau and Lifshitz a long time ago \cite{LL}.
This contribution is not merely proportional to the external magnetic field because of the influence of
other processes with a field- independent rate, e.g. such as the  relaxation of spin polarization or an additional 
spin polarization induced by scattering of polarized magnetic moments.

Despite of the numerous studies of the HE,  the possibility of the ISHE contribution in a ferromagnetic metal was never addressed experimentally or theoretically. In contrast, the ISHE contribution to the HE in a paramagnetic material has been verified experimentally. In equilibrium the electron gas is not spin- polarized in a non-magnetic material and there is no ISHE contribution. Only when the spin polarization is externally created, the ISHE contribution can be detected and identified. For example, the spin polarization in a paramagnetic AlGaAs/GaAs heterojunction was created by circularly- polarized light. The dependence of the measured Hall angle on the degree of circular polarization and therefore on the spin polarization is clearly detected \cite{r8} confirming the existance of the ISHE contribution to HE. Additional proof of the ISHE contribution is that the measured Hall angle exponentially decreases with an increased distance between the focused spot of a laser beam and the Hall probe as would be expected for a diffused spin current.
 
A measurement of the ISHE contribution in a ferromagnetic metal is more difficult, because of the existence of the ISHE contribution even in equilibrium. Presence of the ISHE can be evaluated from the non-linear dependence of the Hall angle on an external magnetic field.
Another difficulty to detect the ISHE lies in the fact that the average local magnetic moment can be 
field- dependent too, usually due to reordering of magnetic domains, causing nonlinearity in the AHE contribution.   
For example, one cannot pin down the ISHE if the alignment of localized moments occurs 
due to an external magnetic field as  it happens in a paramagnet \cite{Giovannini,Maryenko}.
An experimental setup necessary to identify a pure contribution of the ISHE requires appreciable
independence of the local magnetic moment on the external magnetic field $H$.    
In this sense a nanomagnet made of a ferromagnetic metal with the perpendicular magnetic anisotropy 
(PMA) \cite{PMA-review} can be singled out as a unique object for such a measurement.  In this case the magnetic moments are firmly aligned perpendicularly  to the film surface due to the strong PMA effect. The nano-size of the nanomagnet ensures a one-domain state, in which all localized moments are aligned in one direction.
As a result, the contribution of the AHE becomes essentially independent on the external magnetic field H
and the joint contribution of the OHE and AHE is strictly a sum of field independent 
AHE $\alpha_{\mbox{\scriptsize AHE}}$  and linear OHE  $\sim \alpha_{\mbox{\scriptsize OHE}} H$
terms.   
Such a simple dependence gives an unique possibility to detect this
phenomenon unambiguously, because the  deviation of the field dependence of $\alpha_{\mbox{\scriptsize AHE}}$ from 
the "constant plus linear" dependence can occur solely due to the ISHE.
 
In the present paper we consider a phenomenological theory of the ISHE dependence on the external magnetic 
field $H$ and perform field dependent measurements of the HE angle in  numerous nanomagnets made of 
FeB and Fe$_{0.4}$Co$_{0.4}$B$_{0.2}$ which are the ferromagnetic metals with strong PMA effect.
Comparison of the theoretical and experimental dependence of the Hall angle on an external 
magnetic field provides an unambiguous proof of the importance of the ISHE in a metallic ferromagnet.   

\begin{figure}[b]
\begin{center}
\includegraphics[width=8cm]{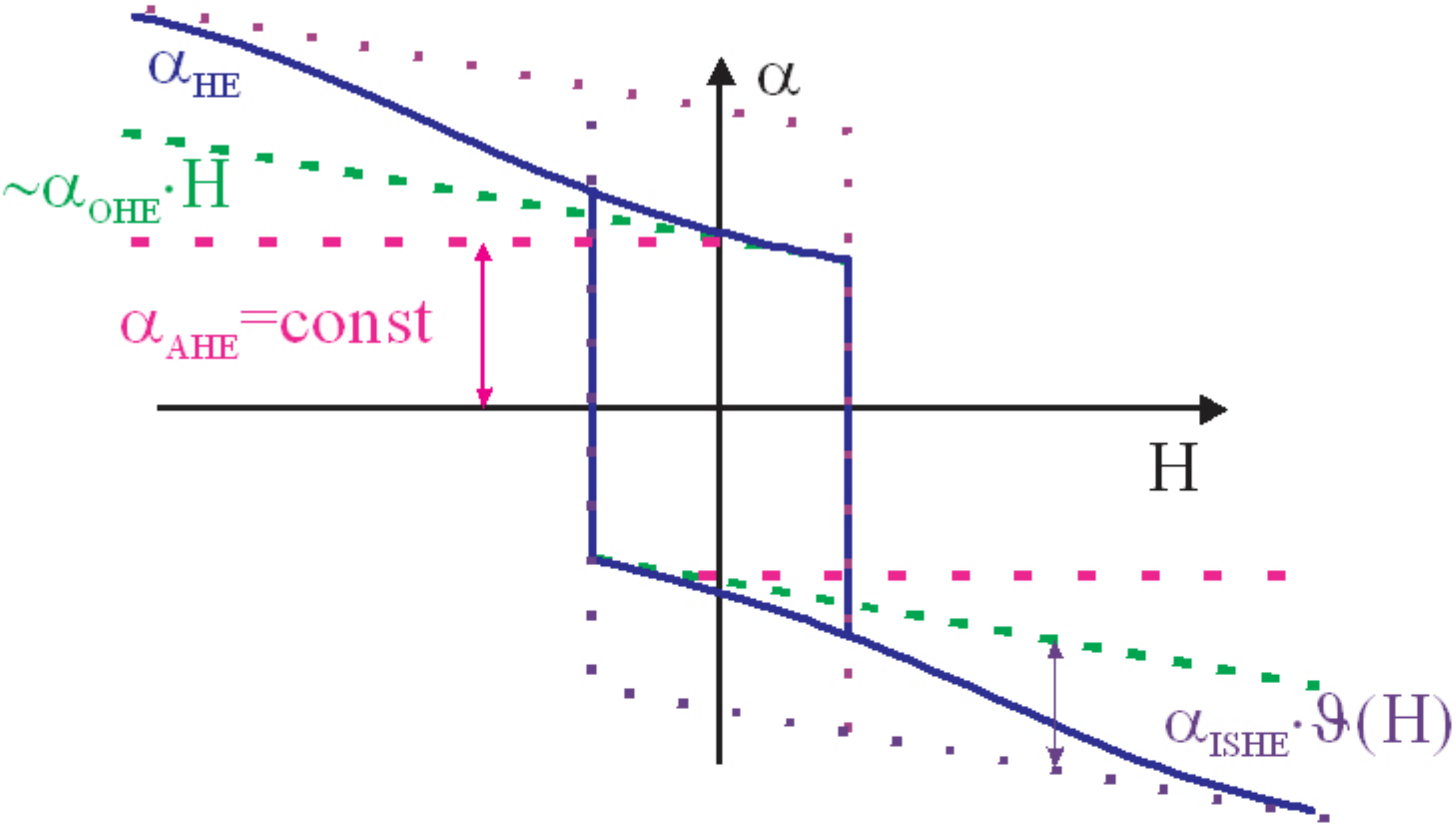}
\end{center}
\caption{\label{fig:fig1} (color online)
Hall angle vs perpendicular magnetic field H. Schematic decomposition of different contributions into Hall angle $\alpha_{HE}$.
Solid blue line is  the measured angle $\alpha_{\mbox{\scriptsize HE}}(H)$.
Dashed (dotted) line is a hypothetical Hall angle in case when ISHE is fixed at 
zero (high) external magnetic field. 
} 
\end{figure}

{\it Theoretical.} 
The Hall angle  $\alpha_{\mbox{\scriptsize HE}}$ in our measurement can be presented as a sum of three
terms (see Fig.)
\begin{equation}
\alpha_{\mbox{\scriptsize HE}} (H)= 
\alpha_{\mbox{\scriptsize OHE}} (\frac{H}{H_0}) +
\alpha_{\mbox{\scriptsize AHE}} +
\alpha_{\mbox{\scriptsize ISHE}} \Theta_{H_S,P_S} (H) \; ,
\label{ha}
\end{equation}
where the scaling field $H_0=1$KGs is introduced to make the coefficient of OHE 
$\alpha_{\mbox{\scriptsize OHE}}$ dimensionless.
The dimensionless coefficien $\alpha_{\mbox{\scriptsize AHE}}$ in our experiment can be 
considered as field independent and
the function $\Theta_{H_S,P_S} (H)$ is defined as a ratio of the contribution of the ISHE at a
finite $H$ and absence of the external magnetic field. 
Figure~\ref{fig:fig1}  shows a schematic representation of these three contributions.
The AHE onset  $\alpha_{\mbox{\scriptsize AHE}}=const$ (horizontal short dash line) 
is summed with linear in $H$ OHE contribution  $\sim \alpha_{\mbox{\scriptsize OHE}} H$ 
giving together the slanted dashed line in Fig.~\ref{fig:fig1}. 
The contribution of the ISHE $\alpha_{\mbox{\scriptsize ISHE}} \Theta_{H_S,P_S} (H)$ 
is the difference between the solid and dashed lines, which is confined between the slanted dotted and dashed lines.

From the theoretical point of view the ISHE originates from the spin-dependent scatterings of the 
conduction electrons \cite{r13,r15,r16,r17,r19,r20,r21,r22}. E.g. in the case of  the scew- scattering mechanism  \cite{r30}, the amounts of spin-polarised conduction electrons scattered into the left/ right directions are different
leading to the ISHE which is, naturally, linearly proportional to the number of spin-polarized electrons.
Since the Hall angle is a ratio of the nondiagonal $\sigma_{xy}$ and diagonal $\sigma_{xx}$ conductivities, 
the ISHE contribution to the Hall angle is proportional to the relative spin polarization   
\begin{equation}
P_s(H) = n_{\mbox{\scriptsize SP}} / \left(  n_{\mbox{\scriptsize SP}}+n_{\mbox{\scriptsize SU}} \right) \; 
\label{p_s}
\end{equation}
in the external magnetic field $H$.
Here, we divided all relevant conduction electrons
$n_{\mbox{\scriptsize SP}}+n_{\mbox{\scriptsize SU}}$, which participate in charge transport,  into the 
group of the spin-unpolarized electrons $n_{\mbox{\scriptsize SU}}$, in which the numbers of spin up and 
down states are equal, and the group of the spin-polarized electrons $n_{\mbox{\scriptsize SP}}$, 
whose spin direction is the same.
Both the spin-polarized and spin-unpolarized states $n_{\mbox{\scriptsize SP}}+n_{\mbox{\scriptsize SU}}$
contribute to the diagonal component $\sigma_{xx}$ of the conductivity whereas only the spin-polarized 
states $n_{\mbox{\scriptsize SP}}$ contribute to $\sigma_{xy}$.       

The conversion rate between spin-polarised 
$n_{\mbox{\scriptsize SP}}$ and spin-unpolarised $n_{\mbox{\scriptsize SU}}$  
electrons can be calculated as \cite{Za2} 
\begin{equation}
\frac{\partial n_{\mbox{\scriptsize SP}}}{\partial t}  =
\left[
\frac{n_{\mbox{\scriptsize SU}}}{\tau_{M_{\perp}}} - 
\frac{n_{\mbox{\scriptsize SP}}}{\tau_{\mbox{\scriptsize rel}}}
\right] 
+
\frac{n_{\mbox{\scriptsize SU}}}{\tau_{H}}
\; ,
\label{d_n}
\end{equation}
where $1/\tau_{M_{\perp}}$ and $1/\tau_{H}$ are the rates of spin-pumping processes 
$n_{\mbox{\scriptsize SU}} \to n_{\mbox{\scriptsize SP}}$ whereas the rate $1/\tau_{\mbox{\scriptsize rel}}$
describes the spin-relaxation  $n_{\mbox{\scriptsize SP}} \to n_{\mbox{\scriptsize SU}}$
into spin unpolarized states.  
The spin pumping rate $1/\tau_{M_{\perp}}$ is set by spin-dependent scattering on the alligned 
magnetic moments $M_{\perp}$ and $1/\tau_{H}$ describes the rate of the spin-polarization processes caused
by the external magnetic field.  
The magnetic field alignment rate in the lowest order is proportional to the external magnetic field $H$, 
$1/\tau_{H}=\varepsilon H$, where $\varepsilon$ is the rate per magnetic field unit.  
The process of the spin alignment is described by the damping term of the Landau-Lifshitz equation \cite{LL}, which is linearly proportional to the magnetic field.  
Recent rigorous calculation of the conversion rate \cite{Za2} led to the same result.

The expression for $\Theta_{H_S,P_S} (H)$ follows from
Eqs.~({\ref{p_s}-\ref{d_n}}) and balance condition $\partial n_{\mbox{\scriptsize SP}}/\partial t=0$,
\begin{equation}
\Theta_{H_S,P_S} (H) = 
\frac
{1+ (H/H_S) / P_s  }
{1+(H/H_S)  }
\; ,
\label{p_s_h2}
\end{equation}
where the spin-polarization in the absence of the external magnetic filed is
$P_s =  \tau_{\mbox{\scriptsize rel}} / 
\left(  \tau_{\mbox{\scriptsize rel}} + \tau_{M_{\perp}}\right)$
and $H_s$ is the scaling magnetic field
$H_S =  1 / \left( \varepsilon \tau_{\mbox{\scriptsize rel}} \right)$
which is determined by the relation between the spin-depolarization relaxation rate 
$1 / \tau_{\mbox{\scriptsize rel}}$ and magnetic field alignment rate per field unit $\varepsilon$ \cite{Zay}. 

\begin{figure}[t]
\begin{center}
\includegraphics[width=8cm]{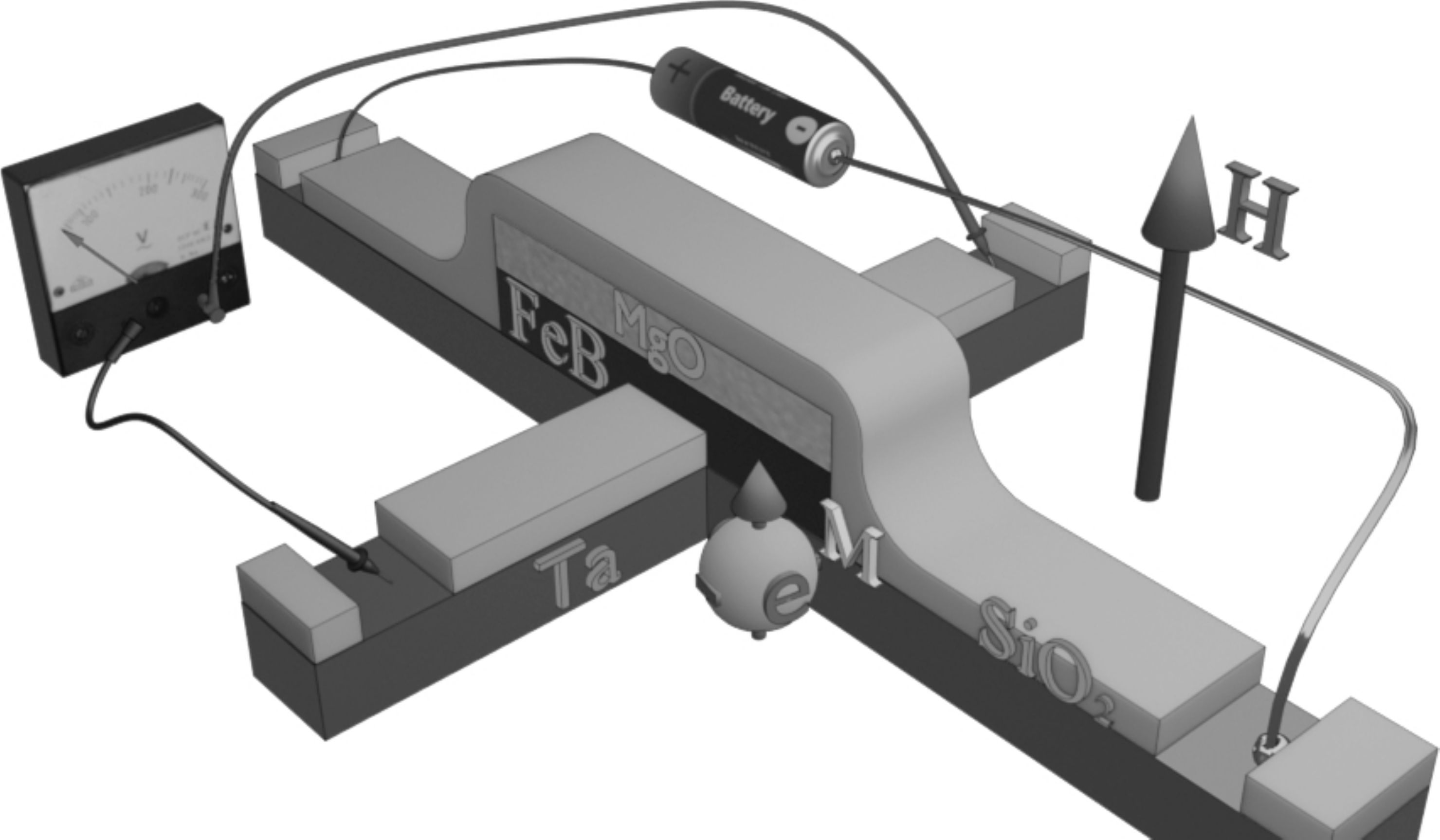}
\end{center}
\caption{\label{fig:fig2} 
Measurement setup. A FeB(1) or Fe$_{0.4}$Co0.4B$_{0.2}$ (1) nanomagnet was fabricated on top of a Ta (3) non-magnetic nanowire. The numbers in parentheses are layer thickness in nm.  The sizes of the nanomagnet are 400 nm x 400 nm. The Hall voltage was measured by a pair of Hall probe aligned to the nanomagnet position.  An external magnetic field H applied along the magnetization direction and perpendicularly to nanomagnet surface. It is critically important for this measurement that the magnetization M is perfectly aligned to the direction of H and there are no magnetic domains in the whole scan range of H.
} 
\end{figure}

{\it Experimental}. The Hall angle $ \alpha_{\mbox{\scriptsize HE}}$ 
was evaluated from the measured Hall voltage using the standard relation \cite{measurement}. The Hall voltage is measured  by the "Hall bar"  measurement setup 
shown in Fig.~\ref{fig:fig2}.
All terms in (\ref{ha}) reverse their sign, when $M$ and $H$ are reversed. 
In order to avoid a systematic error due to a possible misalignment of the Hall probe, the Hall angle was measured as $\alpha = \left[ \alpha(H,M) - \alpha(H,-M) \right] / 2$. Additionally, this allows to exlude all componenets of $\alpha$, which are  independent of the polarity of the magnetic field, such as  $\sim  H^2,  M^2,  M\cdot H$.

\begin{figure}[h]
\begin{center}
\includegraphics[width=8cm]{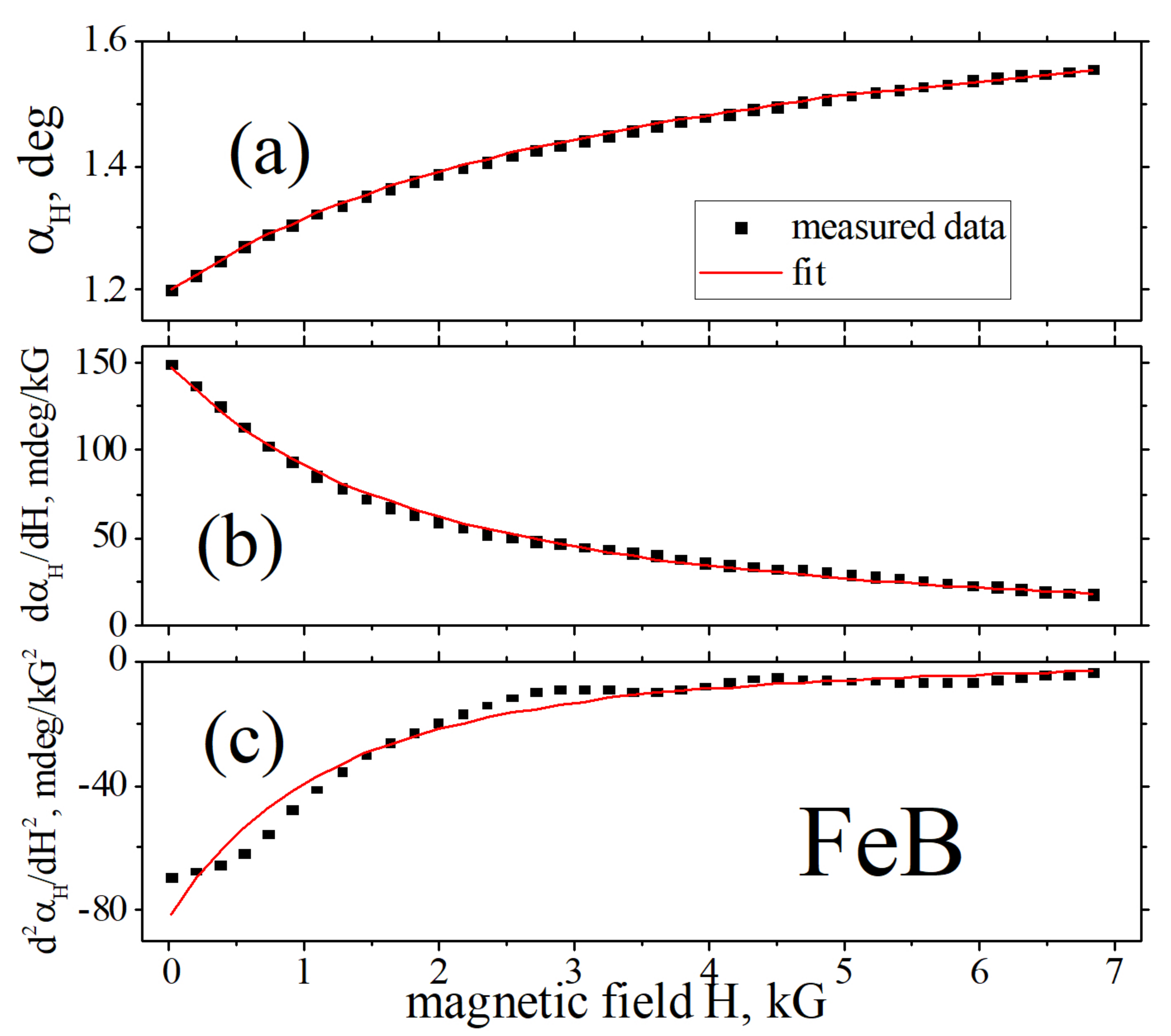}
\end{center}
\caption{\label{fig:fig3} (color online)
Comparison of experimental data (circles) and theoretical fit (lines) for FeB: 
(a) Hall angle $\alpha_{\mbox{\scriptsize Hall}}$ as a  function of the applied magnetic field H; 
(b) first derivative, and (c) second derivative of $\alpha_{\mbox{\scriptsize Hall}}$.
The fit parameters are $\alpha_{\mbox{\scriptsize OHE}}=0.2$mdeg, $H_S=3.954$kG, $P_S=0.5$,
$\alpha_{\mbox{\scriptsize AHE}}=675$mdeg, and $\alpha_{\mbox{\scriptsize ISHE}}=551$mdeg.
}
\end{figure}

{\it Comparison with experiment}.
The measured variation of $\alpha(H)$ vs. the external magnetic field $H$ 
can be explained in terms of the relation (\ref{ha}) depending on five fit parameters,
namely $\alpha_{\mbox{\scriptsize OHE}}$, $\alpha_{\mbox{\scriptsize AHE}}$, 
$\alpha_{\mbox{\scriptsize ISHE}}$, $H_S$, and $P_S$. 
Figure 3 shows that the fit of the experimental data with the above relation is excellent proving the 
importance of the ISHE contribution to the HE in a ferromagnetic metal. 
Although, the obtained fit of experimental data is ambiguous for the shown parameters, we can show that the 
coefficient $\alpha_{\mbox{\scriptsize ISHE}} \ne 0$ and, hence, our data provides
doubtless indication of the ISHE contribution.

Even though the fitting ambiguity does not allow unambiguous evaluation of 
$\alpha_{\mbox{\scriptsize AHE}}$, $\alpha_{\mbox{\scriptsize ISHE}}$,  $P_S$, 
it is possible to trace the tendency of their change  as the parameters of the nanomagnet are 
changed (e.g. temperature, bias current etc.). 
The ambiguity is originated from the fact that the functional dependence (\ref{ha}) does not change when the set of three initial fitting
parameters 
$\{ \alpha_{\mbox{\scriptsize AHE}}, \alpha_{\mbox{\scriptsize ISHE}},  P_S \}$
change to a new set
$\{ \alpha_{\mbox{\scriptsize AHE}}', \alpha_{\mbox{\scriptsize ISHE}}', P_S' \}$
which is related to the initial set as
\begin{equation}
\alpha_{\mbox{\scriptsize ISHE}}'  = 
\alpha_{\mbox{\scriptsize ISHE}} (1-P_S) P_S' / (1-P_S') /P_S \;,
\label{sc2}
\end{equation}
and 
\begin{equation}
\alpha_{\mbox{\scriptsize AHE}}'  = \alpha_{\mbox{\scriptsize AHE}} + 
\alpha_{\mbox{\scriptsize ISHE}} (P_S-P_S') / P_S (1-P_S') \; .
\label{sc3}
\end{equation}
The fit in Fig.~(\ref{fig:fig3}) is obtained for one possible set of parameters shown in the figure 
caption. 
The fit gives an unambiguous evaluation of $\alpha_{\mbox{\scriptsize OHE}}=0.2 $mdeg and $H_S=3.954 $kG,
However, one obtains an identical fit for $P_S=0.6/0.4/0.3/0.2$ at
$\alpha_{\mbox{\scriptsize AHE}}=400/859/990/1088$ mdeg and 
$\alpha_{\mbox{\scriptsize ISHE}}=827/367/236/137$ mdeg. 
Note, the decrease of $\alpha_{\mbox{\scriptsize AHE}}$ is correlated with the increase of 
$\alpha_{\mbox{\scriptsize ISHE}}$.
A similar set of parameters is obtained for field dependence of Hall angle in Fe$_{0.4}$Co$_{0.4}$B$_{0.2}$:
$\alpha_{\mbox{\scriptsize OHE}}=0.2$ mdeg, $H_S=8.8$ kG, $P_S=0.5$,
$\alpha_{\mbox{\scriptsize AHE}}=1497$ mdeg, and $\alpha_{\mbox{\scriptsize ISHE}}=377$ mdeg.

Note, regardless of the ambiguity of the fitting constants, the experimental first derivative 
of $\alpha_{\mbox{\scriptsize HE}} (H)$ in Fig.~(\ref{fig:fig3}) is not constant and the second 
derivative is non- zero. This implies that the non-linear theoretical function $\Theta_{H_S,P_S} (H)$, 
which describes the ISHE, has a non- zero contribution to the full Hall angle (\ref{ha}). 
Additionally, the fitting is only possible when  $P_S \ne 1$ and $P_S \ne 0$. 
The condition $P_s \ne 0$ and Eq. (\ref{sc2}) imply that $\alpha_{\mbox{\scriptsize ISHE}}'$ 
never turns to zero at any set of fitting parameters. 
  
\begin{figure}[t]
\begin{center}
{\includegraphics[width=8cm]{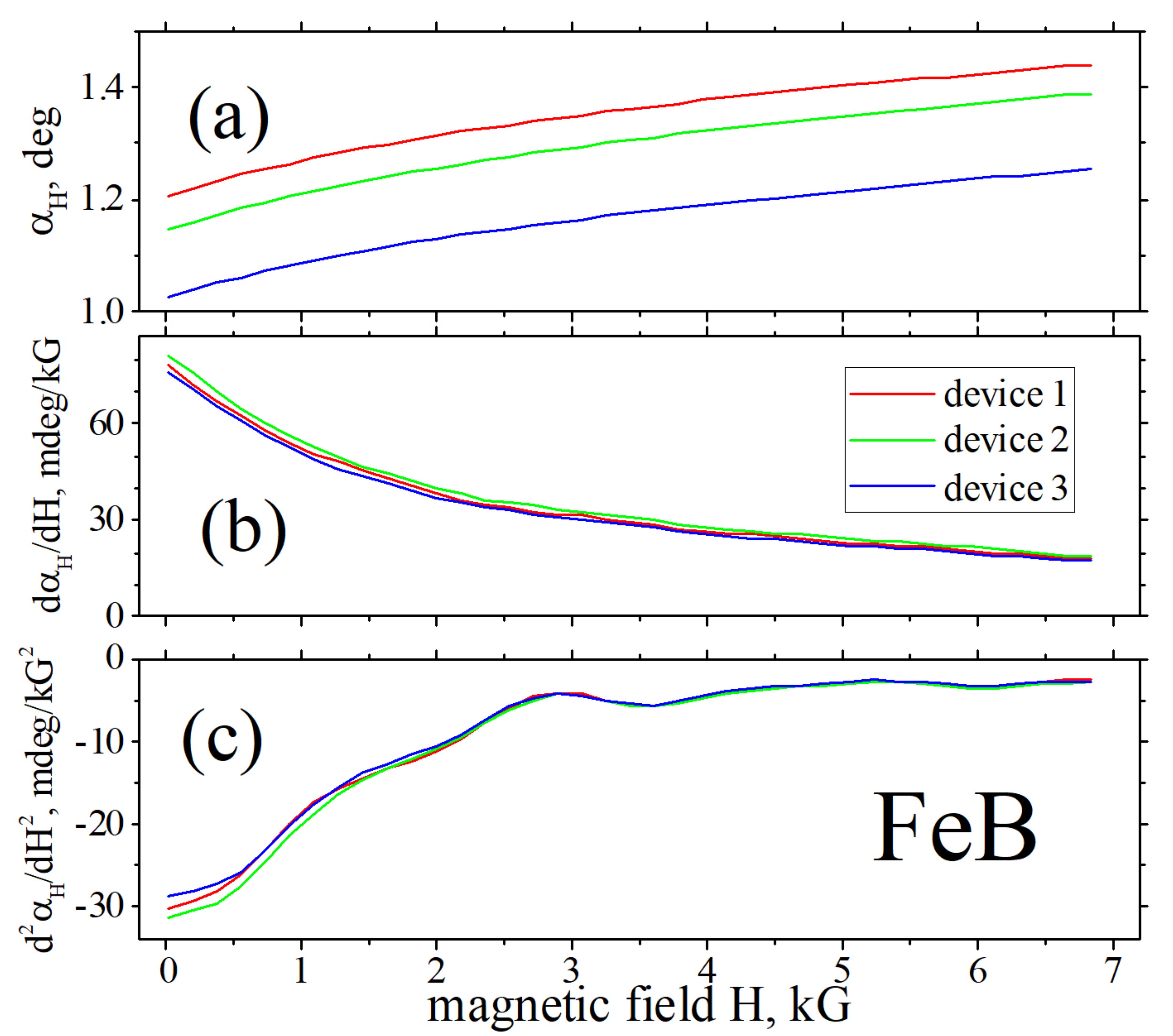}}
\end{center}
\caption{\label{fig:fig4} (color online) 
Hall angle in similar devices (nanomagnets), which are fibricated in different parts of the same wafer.
} 
\end{figure}

\begin{figure}[t]
\begin{center}
{\includegraphics[width=8cm]{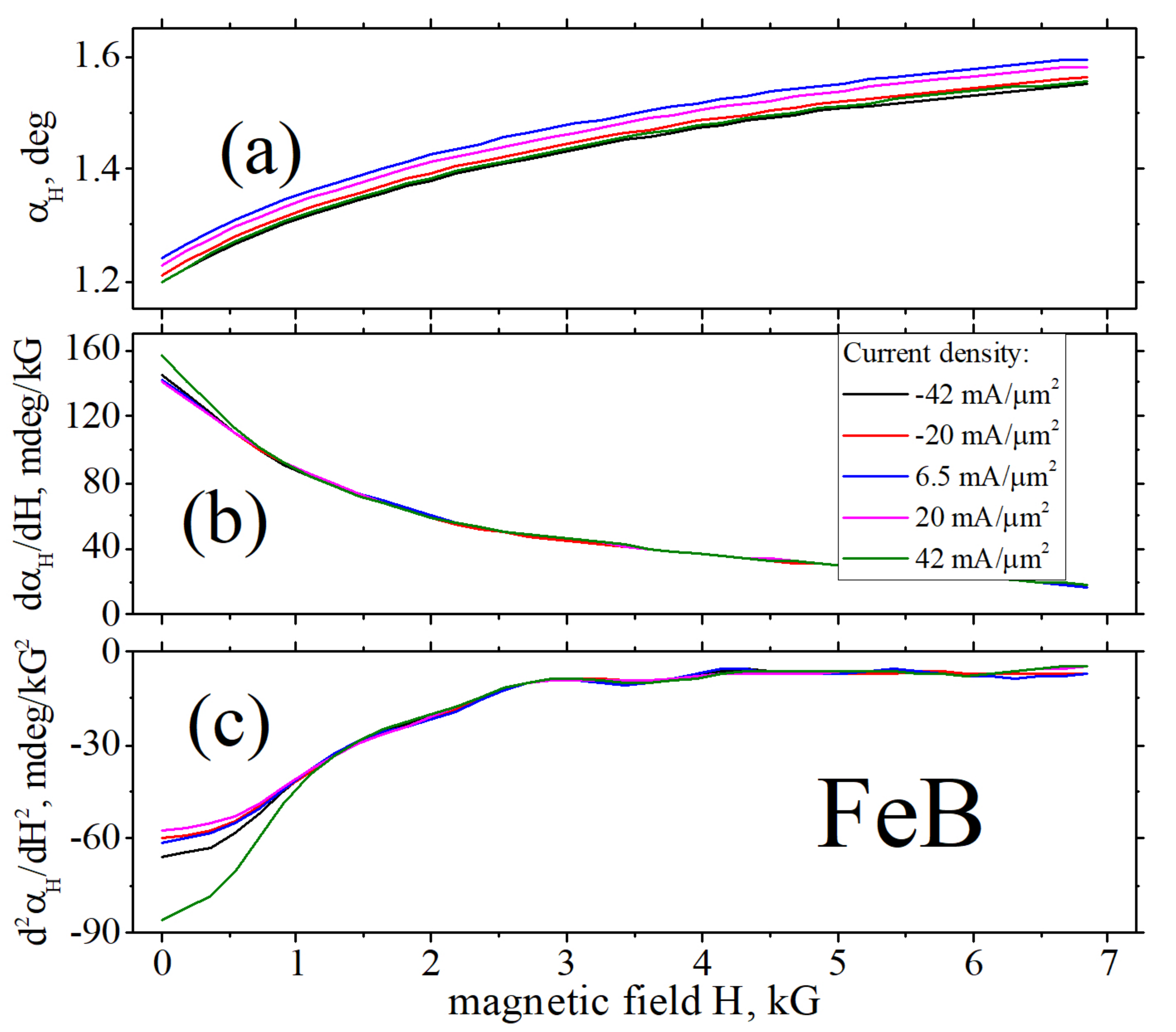}}
\end{center}
\caption{\label{fig:fig5} (color online)
Hall angle measured in the same device at  different current densities. 
} 
\end{figure}

 In order to prove that the magnetic moment of the studied ferromagnetic nanomagnets with a strong PMA is in the single- domain state and is not realigned when an external magnetic field is applied along it,
 we performed the measurement of HE in multiple devices at different 
current densities leading to a slight temperature change.
A typical comparison for different  devices and temperatures is given in Fig.~\ref{fig:fig4} and 
Fig.~\ref{fig:fig5}, respectively. 
Note, the first and second derivatives remain nearly the same showing that the 
OHE and ISHE contributions are nearly the same in different samples and their temperature  dependence is weak.
This nearly- identical dependence of derivatives in different nanomagnets excludes the possibility of existence of any magnetic domains, because the movement of the domain wall should be individual and different for each nanomagnet due to different distributions of fabrication defects and edge irregularities and a slight difference in shape of the nanomagnets.

It should be noted that the firm alignment of local magnetic moments along the magnetic field is the critical requirement for the described measurements. This requirement makes it difficult to perform a similar measurement in an antiferromagnetic, compensated ferromagnetic or paramagnetic material. Additionally, it requires a measurement of the second derivative of $\alpha_{\mbox{\scriptsize H}}$  with a reasonably high signal- to- noise ratio. It is only possible in a material having a low $\alpha_{\mbox{\scriptsize OHE}}$ and a long spin relaxation time. The amorphous FeB and FeCoB are well fit to all these requirements. 

We measured the Hall effect in a ferromagnetic nanomagnet with a strong perpendicular magnetic 
anisotropy, in which  the local magnetic moments are very stable and well- aligned along the easy axis of the nanomagnet. The magnetization direction does not change under an external magnetic field applied along the magnetization.
The adopted expectation is that the Hall angle in a such nanomagnet must be a sum of a linear term proportional to the
external magnetic field $H$, which is the contribution from the ordinary Hall effect, and a field- independent constant, which is the contribution of the
anomalous Hall effect in a sample with field- independent local moments.   
Our measurement revealed an extra non- linear contribution indicating an additional contribution to the 
Hall effect.
We interpreted this contribution as the inverse spin Hall effect, which is originated from the spin-imbalanced scatterings of 
the spin-polarized conduction electrons, and developed a phenomenological theory describing the spin
polarization of conduction electrons of a ferromagnet in an external magnetic field. 
The theoretical expression of our phenomenological approach is in perfect agreement with 
experimentally obtained non- linear contribution giving, thus, a proof of the importance of the  
inverse spin Hall effect in a ferromagnetic metal.

{\it Acknowledgements}. We acknowledge stimulating discussions with N. Nagaosa.
This work was supported by JST CREST Grant Number JPMJCR1874, Japan.


\begin{thebibliography}{99}
%
\bibitem{Hall1879} E. H. Hall,  Am. J. Math. \textbf{2}, 287 (1879).
%
\bibitem{PughRostoker1953} E. M. Pugh and N. Rostoker, Rev. Mod. Phys. \textbf{25}, 151 (1953).
%
\bibitem{Sinova} N. Nagaosa, J. Sinova, S. Onoda A. H. MacDonald, and N. P. Ong,  
                             Rev. Mod. Phys. \textbf{82}, 1539-1592 (2010).
%
\bibitem{Saitoh} E. Saitoh, M. Ueda, H. Miyajima, and G. Tatara, 
                    Appl. Phys. Lett. \textbf{88}, 182509 (2006).
%
\bibitem{Tinkham} S. O. Valenzuela and M. Tinkham, 
                                     Nature \textbf{442}, 176 (2006).
%
\bibitem{Maekawa} T. Kimura, Y. Otani, T. Sato, S. Takahashi, and S. Maekawa, 
Phys. Rev. Lett. \textbf{98}, 156601 (2007).
%
\bibitem{r30} J. Smit, Physica  \textbf{21}, 877 (1955).
%
\bibitem{LL}  L. D. Landau and E. M. Lifshitz,  Phys. Z. Sowjetunion. \textbf{8}, 153 (1935).
%
\bibitem{r8} J. Wunderlich, A. C. Irvine, J. Sinova, B. G. Park, L. P. Z\^{a}rbo, X. L. Xu, B. Kaestner, 
                    V. Nov\'{a}k, and T. Jungwirth, Nat. Phys. \textbf{5}, 675 (2009).
%
\bibitem{Giovannini} B. Giovannini, J. Low Temp. Phys. \textbf{11}, 489 (1973).
%
\bibitem{Maryenko} D. Maryenko, A. S. Mishchenko, M. S. Bahramy, A. Ernst, J. Falson, Y. Kozuka, 
                                 A. Tsukazaki, N. Nagaosa, and M. Kawasaki, Nat. Comm. \textbf{8}, 14777 (2017).
%
\bibitem{PMA-review} M. T. Johnson, P. J. H. Blomen, F. J. A. den Broeder, and J. J. de Vries, 
                    Rep. Prog. Phys. \textbf{59}, 1409 (1996).
%
\bibitem{r13} C. M. Hurd, {\it The Hall Effect in Metals and Alloys} (Plenum Press, 1972).
%
\bibitem{r15} A. Fert and O. Jaoul, Phys. Rev. Lett. \textbf{28}, 303 (1972).
%
\bibitem{r16} A. Fert and A. Friederich, Phys. Rev. B \textbf{13}, 397 (1976).
%
\bibitem{r17} J. Kondo, Prog. Theor. Phys. \textbf{27}, 772 (1962).
%
\bibitem{r19} R. Karplus and J. M. Luttinger, Phys. Rev. \textbf{95}, 1154 (1954).
%
\bibitem{r20} J. Smit, Physica \textbf{24}, 39 (1958).
%
\bibitem{r21} F. E. Maranzana, Phys. Rev. \textbf{160}, 421 (1967).
%
\bibitem{Za2} V. Zayets, J. Magn. Magn. Mater.  \textbf{445},  53 (2018).
%
\bibitem{r22} J. M. Luttinger and W. Kohn, Phys. Rev. \textbf{97}, 869 (1955).
%
\bibitem{Zay} V. Zayets, J. Magn. Magn. Mater.  \textbf{356},  52 (2014). %
%
\bibitem{measurement} 
$\alpha= 
\left[ 
1 + \sigma_{\mbox{\scriptsize nM}} t_{\mbox{\scriptsize nM}}/
\sigma_{\mbox{\scriptsize f}} t_{\mbox{\scriptsize f}}
\right]
V_{\mbox{\scriptsize H}}L /  (IRw)$,
where $\sigma_{\mbox{\scriptsize f}}$  and $\sigma_{\mbox{\scriptsize nM}}$  
are conductivities of ferromagnetic and non-magnetic metals,   
$t_{\mbox{\scriptsize f}}$  and $t_{\mbox{\scriptsize nM}}$ are their thicknesses, 
$V_{\mbox{\scriptsize H}}$ is the measured Hall voltage, 
$I$ is the bias current and $R$, $L$, and $w$ are the resistance, length and width of the nanowire,
correspondingly. 
%
%
%

\end{thebibliography}
\end{document}